\documentclass{article}[a4paper,11points]
\usepackage{graphicx} % Required for inserting images
\usepackage{amsfonts,amsmath,amssymb}
\usepackage{slashed}

\usepackage{dynkin-diagrams}
\usepackage{cite}
\usepackage{amsmath}
\usepackage{amssymb}
\usepackage{amsthm}
\usepackage{mathrsfs}
\usepackage{xypic}
\usepackage{tikz-cd}
\usepackage{chngcntr}
\usepackage{enumitem}
\usepackage{authblk}
\usepackage[normalem]{ulem}

\usepackage{contour}

\contourlength{0.8pt}

\usepackage{hyperref}
\usepackage{color}
\definecolor{darkred}{rgb}{0.65,0.15,0}
\definecolor{ForestGreen}{rgb}{0.302,0.561,18.92}
\hypersetup{pdfborder={0 0 0},colorlinks=true,urlcolor=darkred,citecolor=blue,linkcolor=darkred,linktocpage=true}

\usepackage{float} 

\usepackage{geometry}
\usepackage{cancel}

\newgeometry{vmargin={35mm}, hmargin={35mm}} 

\usepackage{youngtab}
\def\youngdim{8pt}

\Yboxdim\youngdim

\def\eg{{\it e.g.}}

\def\DWeight#1#2#3{\bigl(\raise2.5pt\hbox{${}_{#1}$}{}^{#2}_{#3}\bigr)}

\def\AAWeight#1#2{\bigl(\raise0pt\hbox{${}^{#1}_{#2}$}\bigr)}

\def\so{{\mathfrak{so}}}

\def\sp{{\mathfrak{sp}}}
\def\su{{\mathfrak{su}}}

\def\Spin{\mathrm{Spin}}
\def\SU{\mathrm{SU}}
\def\SL{\mathrm{SL}}
\def\U{\mathrm{U}}

\def\nn{\nonumber}

\def\so{\mathfrak{so}}

\def\*{\partial}

\def\RR{{\mathbb R}}
\def\CC{{\mathbb C}}
\def\HH{{\mathbb H}}
\def\OO{{\mathbb O}}
\def\KK{{\mathbb K}}

\def\Re{\hbox{Re}\,}
\def\Im{\hbox{Im}\,}

\def\ZZ{{\mathbb Z}}

\def\tr{\hbox{tr\,}}

\newlength\symlength
\newlength\pluslength
\DeclareRobustCommand\lplus{
 \ensuremath{
  \mathop{
   \begin{tikzpicture}[line width=0.1ex]
    \useasboundingbox (-1ex, -0.7ex) rectangle (.7ex, 1ex);
\draw (55:\symlength) arc (55:305:\symlength);
\draw (-\pluslength,0) -- (\pluslength,0);
\draw (0,-\pluslength) -- (0,\pluslength);
   \end{tikzpicture}
  }\nolimits
 }
}

\def\inplus{\lplus}

\makeatletter
\DeclareRobustCommand{\cev}[1]{%
  {\mathpalette\do@cev{#1}}%
}
\newcommand{\do@cev}[2]{%
  \vbox{\offinterlineskip
    \sbox\z@{$\m@th#1 x$}%
    \ialign{##\cr
      \hidewidth\reflectbox{$\m@th#1\vec{}\mkern4mu$}\hidewidth\cr
      \noalign{\kern-\ht\z@}
      $\m@th#1#2$\cr
    }%
  }%
}
\makeatother

\def\Cev#1{\overset\leftarrow{#1}}
\def\Vec#1{\overset\rightarrow{#1}}

\makeatletter
\newcommand{\oset}[3][0ex]{%
  \mathrel{\mathop{#3}\limits^{
    \vbox to#1{\kern-3\ex@
    \hbox{$\scriptscriptstyle#2$}\vss}}}}
\makeatother

\def\slash#1{#1\mkern-7.5mu/}

\title{Infinite spin superparticles\vspace{24pt}}

\author[1]{Eugenia Boffo\thanks{boffo@karlin.mff.cuni.cz }}
\author[2]{Martin Cederwall\thanks{martin.cederwall@chalmers.se}\vspace{12pt}}

\affil[1]{\small{Faculty of Mathematics and Physics, Mathematical Institute, Charles University Prague, Sokolovsk\'a~83, 186 75 Prague, Czech Republic}\vspace{6pt}}
\affil[2]{\small{Dept. of Physics, Chalmers University of Technology, 412 96 Gothenburg, Sweden}\vspace{6pt}}

\date{{\small November, 2026}\vspace{24pt}}

\numberwithin{equation}{section}

\begin{document}

\frenchspacing

\clearpage\maketitle
\thispagestyle{empty}

%\maketitle

\begin{abstract}
\noindent We obtain infinite spin representations of the super-Poincar\'e algebra from a higher-derivative superparticle action on a super-Minkowski background in 3, 4, 6 and 10 dimensions.
Super-twistor transforms are performed, and used to deduce first-quantised spectra. 
\end{abstract}

\newpage

\tableofcontents

\newpage

\section{Introduction}

Higher spin particles \cite{Wigner:1939cj} have long been under inspection for the formulation of a theory of quantum gravity. String theory is renowned to have higher spins, see for instance \cite{Sagnotti:2011jdy} for a review of the interplay between strings and higher spin particles. 
The development of higher spin theories is far from complete, as they face some problems: for example interacting theories combining spins greater than two are difficult to construct, though there is constant  progress, see the recent ref.~\cite{Basile_2023}. 

Similarly in spirit, another class of higher spin particles was observed even further in the past, in Wigner's classification of particles \cite{Wigner:1939cj}, based on mass and spin. He noticed that the norm squared of the Pauli--Luba\'{n}ski vector could also be non-zero. This is the starting point for \emph{infinite} or \emph{continuous spin} particles. 
The adjective ``continuous'' is rather deceiving, since the value of the spin is still discrete (integer and half-integer). Although still the dominating terminology,
we prefer to
use the term ``infinite spin''. 
Wigner soon dismissed them because a gas of these particles would have infinite heat capacity \cite{Wigner1997}, a property that has however later been disproven \cite{Schuster:2024wjc}. 
Infinite spin particles have never been observed in nature, and are not even an output of perturbative string theory \cite{Font:2013hia,Alabbasi:2026kbf}. Putting these possible drawbacks aside, the fascination on the subject still persists: a recent example is \cite{Basile:2026ntj}, where the spinor-helicity formalism is developed for infinite spin particles.

In the following we will present a derivation of target space supersymmetric infinite spin particles from an action functional with higher world-line derivatives (it depends on the acceleration). 
Supersymmetry is another vital concept of modern theoretical physics, as it helps towards combining gravity and quantum field theory. Therefore it is natural to ask about a supersymmetric  extension of the infinite spin particles, which would arrange spins in super-multiplets (representations of supersymmetry). 
After setting down the formalism and defining the dynamics of the super-particles,
we will move on to do a super-twistor transform on our phase space, motivated by the fact that the Hamiltonian analysis of constraints has produced a mixture of first and second class constraints. In fact, the model we will be studying bears very strong resemblance to the Brink--Schwarz superparticle model, which we will comment on. In the last part we will provide a quantisation, 
yielding natural super-multiplets. Our considerations span the dimensions $3, 4, 6, 10$ --- 2 higher that the dimensions of the division algebras --- and provide super-Poincar\'e covariant versions and generalisations of the light-c\^one models of \cite{Brink:2002zx}. 

In $D=10$, where local symmetries form $L_\infty$ algebroids, with $S^7$ as an ingredient, we provide BRST operators for precise resolutions.

We will refrain from presenting a gauge field theory of infinite spins.    
For an action principle, see Schuster and Toro \cite{Schuster:2013pta} for the bosonic version, and \cite{Bekaert:2015qkt} for the fermionic, as well as \cite{Najafizadeh:2019mun} for an explicit supersymmetric version. All our considerations are in flat (Minkowski) spacetime; for the curved case, 
see for example
\cite{Bekaert:2023uve,Anninos:2025mje,Buchbinder:2024hea,Buchbinder:2025gsr}
among other references.

\section{Infinite spin particles on superspace\label{SuperParticleSection}}

In an ingenious paper \cite{Edgren:2005gq}, Edgren, Marnelius and Salomonson derived infinite spin representations from particle dynamics defined by a higher derivative action.
The extra phase space variables, compared to an ordinary massless particle, were obtained through the Ostrogradsky formalism \cite{Ostrogradsky:1850fid}, applied to an action containing a term proportional to the acceleration.  
The present construction is a direct (target space) supersymmetric generalisation of that work. 

\subsection{Action}

The construction we undertake is based on a bosonic world-line and a supersymmetric target space, like the Brink--Schwarz superparticle \cite{Brink:1981nb,Bengtsson:1984rw}. We do not impose any world-line supersymmetry, present for so-called spinning particles \cite{Brink:1976sz,Brink:1976uf,Boffo:2024lwd,Boffo:2025fip}.
Our dynamical objects thus consist of a map from a (bosonic) world-line, parametrised by a coordinate $t$, to a $(d|s)$-dimensional (flat) superspace with coordinates $Z^M=(x^m,\theta^\mu)$.
The superspace vielbein is 
\begin{align}
E_M{}^A
=\left(\begin{matrix}
    E_m{}^a&E_m{}^\alpha\\E_\mu{}^a&E_\mu{}^\alpha
\end{matrix}\right)
=\left(\begin{matrix}
    \delta_m{}^a&0\\-(\gamma^a\theta)_\mu&\delta_\mu{}^\alpha
\end{matrix}\right)\;.
\end{align}
We denote the world-line einbein by $e$, this is a world-line density.

Define the accelerations
\begin{align}
    Y^a={d\over dt}\left(e^{-1}\dot Z^ME_M{}^a\right)={d\over dt}\left(e^{-1}(\dot x^a+(\theta\gamma^a\dot\theta))\right)\;.
\end{align}
The $Y^a$'s are scalar densities. We will apply the Ostrogradsky formalism (see Appendix \ref{OstroApp} for conventions in the graded case) to the action
\begin{align}
S=\int dt\left(\sqrt{Y^2}+e\mu\right)\;, \quad {{\mu \in \mathbb{R}}}.
\end{align}

\subsection{Constraint analysis}

Let $y^a=\dot x^a$ with momentum $r_a$. Denote the momentum to $\theta^\alpha$ by $\pi_\alpha$ and the momentum to $\eta^\alpha=\dot\theta^\alpha$ by $\varrho_\alpha$. The momenta are
\begin{align}
p_a&=-{1\over\sqrt{Y^2}}\left(\delta_a^b-{Y_aY^b\over Y^2}\right)\dot Y_b\;,\nn\\
r_a&={1\over\sqrt{Y^2}}e^{-1}Y_a\;,\nn\\
\pi_\alpha&=-p_a(\gamma^a\theta)_\alpha+r_a(\gamma^a\eta)_\alpha\;,\\
\varrho_\alpha&=-r_a(\gamma^a\theta)_\alpha\;,\nn\\
p_e&=-e^{-1}(y^ar_a+\eta^\alpha\varrho_\alpha)\;.\nn
\end{align}
We immediately obtain the primary constraints
\begin{align}
T_1&=p\cdot r\approx0\;,\nn\\
T_2&=r^2-e^{-2}\approx0\;,\nn\\
\tau_{1\alpha}&=\pi_\alpha+p_a(\gamma^a\theta)_\alpha -r_a(\gamma^a\eta)_\alpha\approx0\;,\\
\tau_{2\alpha}&=\varrho_\alpha +r_a(\gamma^a\theta)_\alpha\approx0\;,\nn\\
w&=ep_e+y^ar_a+\eta^\alpha\varrho_\alpha\approx0\;.
\end{align}
The Hamiltonian is $H=\dot ep_e+\dot x^ap_a+\dot y^ar_a
+\dot \theta^\alpha\pi_\alpha+\dot\eta^\alpha\varrho_\alpha-L$.
It is straightforward to show that
$\dot ep_e+\dot y^ar_a+\dot\eta^\alpha\varrho_\alpha
=\sqrt{Y^2}=L-e\mu$, so
\begin{align}
H=y^ap_a+\eta^\alpha\pi_\alpha-e\mu\;.
\end{align}
Secondary constraints are obtained by time evolution:
\begin{align}
\{H,T_1\}&=-p^2\;,\nn\\
\{H,T_2\}&=-2T_1\;,\nn\\
\{H,\tau_{1\alpha}\}&=2p_a(\gamma^a\eta)_\alpha\;,\\
\{H,\tau_{2\alpha}\}&=-\tau_{1\alpha}\;,\nn\\
\{H,w\}&=-H\;.\nn
\end{align}
The new constraints are $T_0=p^2\approx0$, $H\approx0$ and $\tau_{0\alpha}=p_a(\gamma^a\eta)_\alpha\approx0$.
No further constraints are generated.

Note that the constraint $w$
is a grading operator counting degree of homogeneity in $e,y,\eta$ (minus their conjugates). 
This is identical to the scaling weight under diffeomorphisms ($x,\theta$ are scalars, while $e,y,\eta$ are densities of weight 1). 
All constraints have definite weight, so $w$ is first class and can be used to gauge fix\footnote{In ref. \cite{Edgren:2005gq}, it is stated that $e$ becomes dynamical. We do not reproduce that result. World-line diffeomorphism invariance guarantees it is pure gauge.} to $e=1$.

All the remaining even constraints ($T_{0,1,2}$ and $H$) are first class. The non-vanishing Poisson brackets between odd constraints are
\begin{align}
\{\tau_{0\alpha},\tau_{2\beta}\}
    &=p_a\gamma^a_{\alpha\beta}\;,\nn\\
\{\tau_{1\alpha},\tau_{1\beta}\}
    &=2p_a\gamma^a_{\alpha\beta}\;.
\end{align}
$\tau_1$ contains the same mixture of first and second class constraints \cite{Bengtsson:1984rw} as the spinorial constraint of the Brink--Schwarz superparticle
\cite{Brink:1981nb}. 
We need to examine the content of $\tau_0$ and $\tau_2$.
Given that $p^2=0$, $\tau_0$ itself only effectively contains half a spinor, which has a Poisson bracket of full rank with half of $\tau_2$. These two half spinor constraints are second class, while the remaining half of $\tau_2$ is first class. 
This can be seen explicitly by choosing a frame where 
\begin{align}
\slash p=\left(\begin{matrix}1&0\\0&0\end{matrix}\right)\;,\qquad
\slash r=\left(\begin{matrix}0&1\\1&0\end{matrix}\right)\;
\end{align}
(times some irrelevant numbers).
Then the fermionic constraints consist of the five $\so(d-2)$ spinors
\begin{align}
(1)\qquad0&\approx\eta^1\;,\nn\\
(2)\qquad0&\approx\pi_1+\theta^1-\eta^2\;,\nn\\
(3)\qquad0&\approx\pi_2-\eta^1\;,\\
(4)\qquad0&\approx\varrho_1+\theta^2\;,\nn\\
(5)\qquad0&\approx\varrho_2+\theta^1\;.\nn
\end{align}
The constraints (1) and (4) are second class.
The first class constraint (5) may be used to gauge fix 
$\eta^2$ to 0.
When this gauge is chosen, $\eta=0$, which is second class together with $\tau_2$, and $\eta$ and $\varrho$ are eliminated. No correction terms in Dirac brackets arise for the remaining variables. 
The remaining odd constraints ((2), second class, and (3), first class) become identical to those of the Brink--Schwarz superparticle.

After the above gauge fixing, the phase space variables are
$x^a,y^a,\theta^\alpha$ and their conjugates
$p_a,r_a,\pi_\alpha$, with the constraints
\begin{align}
T_0&=p^2\approx0\;,\nn\\
T_1&=p\cdot r\approx0\;,\nn\\
T_2&=r^2-1\approx0\;,\label{InfSpinConstraints}\\
H&=y\cdot p-\mu\approx0\;,\nn\\
\tau_\alpha&=\pi_\alpha+p_a(\gamma^a\theta)_\alpha\approx0\;.\nn
\end{align}

\section{Casimir operators}

The central elements of the universal enveloping algebra are fundamental for classifying irreps. From Lie theoretical considerations, it is possible to show that the number of Casimirs of $\mathfrak{iso}(d-1,1)$ is $[\frac{d+1}{2}]$ \cite{Chaichian:1981wr}. We wish to explain how many Casimir invariants there are in the supersymmetric case, and check their value in our specific realisation.

Let $Q_\alpha$ be
\begin{equation}
    Q_\alpha:=\pi_\alpha-p_a(\gamma^a\theta)_\alpha\;.
\end{equation}
Then the Pauli--Luba\'nski tensor \cite{Buchbinder:1998twe,Zumino:2004nb,Arvanitakis:2016wdn} 
\begin{align}
W_{abc}=p_{[a}M_{bc]}+\tfrac1{96}(Q\gamma_{abc}Q)
\end{align}
is invariant under adjoint action of $p_a$ and $Q_\alpha$ and moreover it is manifestly a tensor under $M_{ab}$. 
It is shown using $\slashed{p}Q=0$. This implies that $W_{abc}W^{abc}$ is a super-Poincar\'e Casimir.

For the Brink--Schwarz superparticle, one has (in the present conventions)
\begin{align}
M^{ab}&=x^{[a}p^{b]}+\tfrac14(\theta\gamma^{ab}\pi)\;,\nn\\
      &=x^{[a}p^{b]}-\tfrac14p_c(\theta\gamma^{abc}\theta)\;,
      \label{MQExpr}\\
Q_\alpha&=\pi_\alpha-p_a(\gamma^a\theta)_\alpha
    =-2p_a(\gamma^a\theta)_\alpha\;,\nn
\end{align}
and $p^2=0$, $p_a(\gamma^aQ)^\alpha=0$. For the Brink--Schwarz superparticle, $W_{abc}=0$.

For an infinite spin superparticle, after the elimination of $\tau_{0,2}$, the only difference from the above is
$M^{ab}=x^{[a}p^{b]}+y^{[a}r^{b]}+\tfrac14(\theta\gamma^{ab}\pi)$. We obtain
$W_{abc}=p_{[a}y_br_{c]}$, and
\begin{align}
6W_{abc}W^{abc}=-(p\cdot y)^2r^2=-\mu^2\;,
\end{align}
using the constraints \eqref{InfSpinConstraints}.

This completes the Casimirs in dimension 4.
In dimensions greater than 4, instead, there could be higher Casimirs. In 6 dimensions, with $N=1$ supersymmetry, the antisymmetric products of two supercharges 
are $(Q\gamma_{abc}Q)$ and $(Q\gamma_ae_iQ)$, where $e_i$, $i=1,2,3$, are matrices generating the R-symmetry $\su(2)$ and commuting with $\gamma$ matrices.
In terms of representations of $\SU(4)\times\SU(2)$:
$\wedge^2(001)(1)=(002)(0)\oplus(010)(2)$.
The matrices $e_i$ generate the $\SU(2)$ R-symmetry, and obey $[\gamma_a,e_i]=0$, $e_ie_j=-\delta_{ij}+\epsilon_{ijk}e_k$. In 2-component formalism, $e_i$ is represented by right multiplication by the imaginary quaternion $-e_i$. We use 
$(Q\gamma_{abc}Q)={1\over6}\epsilon_{abcdef}(Q\gamma^{def}Q)$ and (consistently with the conventions for superparticles above) 
\begin{align}
[Q_\alpha,Q_\beta]&=-2\gamma^a_{\alpha\beta}p_a\;,\nn\\
[Q_\alpha,M_{ab}]&=\tfrac14(\gamma_{ab}Q)_\alpha\;,\\
[Q_\alpha,R_i]&=(e_iQ)_\alpha\;,\nn
\end{align}
where $R_i$ are the generators of R-symmetry.

Consider, for a massless representation, a combination of the terms
\begin{align}
A&=p_{[a}M_{bc}M_{de]}\;,\nn\\
B&=M_{[ab}(Q\gamma_{cde]}Q)\;,\nn\\
C&=\epsilon_{abcdef}(Q\gamma^fe_iQ)R_i\;,
    \label{ABCDdef}\\
D&=\epsilon_{abcdef}p^fR^2\;.\nn
\end{align}
They are all invariant under $p$ and $R$, the second one thanks to self-duality, which makes its variation under $p$ vanish modulo $p^a\gamma_aQ$ (note that this also tells us that such a term does not work in 10 dimensions). Under $Q$, there seems to be a possibility to get invariance. Terms of the types
$pMQ$ (from $A,B$), $pRQ$ (from $C,D$) and $Q^3$ (from $B,C$) are produced. 
The last one comes in two versions, containing the vector-spinors $(\gamma^{bc}Q)_\alpha(Q\gamma_{abc}Q)$
and $(e_iQ)_\alpha(Q\gamma_ae_iQ)$,
but since $\wedge^3(001)(1)=(011)(1)\oplus(100)(3)$, there is only one (automatically $\gamma$-trace-free) vector-spinor, and the two $(Q^3)_{a\alpha}$ expressions need to be proportional through a Fierz identity. We find
$(\gamma^{bc}Q)_\alpha(Q\gamma_{abc}Q) =-4(e_iQ)_\alpha(Q\gamma_ae_iQ)$.
One obtains 3 equations for the 4 coefficients, and the (massless) invariant is
\begin{equation}
    C_6=\tfrac1{120}W_{abcde}W^{abcde}\;, \quad 
    W\equiv A+\tfrac1{48}B+\tfrac1{960}C+\tfrac1{240}D\;.
\end{equation} 
We have not found this super-Poincar\'{e} Casimir elsewhere in the literature.

All our calculations here are in the classical setting, so that ordering is irrelevant, we just consider functions on phase space. 
If we work in quantum theory (with commutators), all expressions entering the invariants are assumed to be given a graded symmetrised ordering.

We now wish to examine the value of $C_6$ for the infinite spin superparticle. 
Inserting the expressions in eq. \eqref{MQExpr} (there is no need to insert the angular momentum with the contribution from the second order derivative, $y_{[a}r_{b]}$: it is possible to show that the linear terms of $A$ and $B$ in $y_{[a}r_{b]}$ cancel out) and
$R_i=(\theta e_i\pi)=-p^a(\theta\gamma_ae_i\theta)$, we get
\begin{align}
A+\tfrac1{48}B&=-\tfrac1{16}p_{[a}p^fp^g
(\theta\gamma_{bc|f|}\theta)(\theta\gamma_{de]g}\theta)\;,
\nn\\
\tfrac1{960}C+\tfrac1{240}D
&=-\tfrac1{240}\epsilon_{abcdef}p^fp^gp^h
(\theta\gamma_ge_i\theta)(\theta\gamma_he_i\theta)\;.
\label{WFiveCalc}
\end{align}
Dualising one of the $\gamma_{(3)}$'s (remembering that it has the opposite chirality to the one in $(Q\gamma_{(3)}Q)$) and using the $(020)(0)$ Fierz identity (following directly from the cubic Fierz identity above)
$(\theta\gamma_a{}^{cd}\theta)(\theta\gamma_{bcd}\theta) =-4(\theta\gamma_ae_i\theta)(\theta\gamma_be_i\theta)$, we find that the two lines in eq. \eqref{WFiveCalc} cancel each other, so
\begin{align}
W_{abcde}=0\;.
\end{align}

In ref. \cite{Buchbinder:2021bgv}, a description of $D=6$ (bosonic) infinite spin particles in terms of spinorial variables (twistors) is given. There, the value of $C_6$ is unrestricted. 
We comment on the relation between that model and ours in the following Section \ref{sec:fourth}.
Possibly, (super-)particles with even higher derivatives may provide enough phase space variables to yield vanishing higher invariants, \eg\ $C_6$ in $D=6$.

Can something be done in $D=10$? We believe that the invariants of $D=10$ stops with $P^2$ and $(P_{[a}M_{bc]})^2$ because we cannot use self-duality of $\gamma_{(3)}$ like in the $D=6$ case and we do not even have an Ansatz.

\section{Supertwistor transforms}
\label{sec:fourth}

Before delving into the supersymmetric constrained system under investigation, let us discuss the \emph{twistor transform} in the non-susy setting, thus dropping the odd variables and the $\tau_\alpha$ constraints. We will perform the twistor transforms in 3, 4, 6 and 10 dimensions. This requires a bit of background in \emph{division algebras},
see Appendix \ref{DivisionAlgApp}.

\subsection{3, 4 and 6 dimensions}

\subsubsection{Massless particles}

We reserve a separate treatment to the 10 dimensional case. Hereinafter, $\nu =\{1,2,4\}$. The ordinary massless twistor transform \cite{Bengtsson:1987si} solves $p^2=0$ in terms of a spinor bilinear in $\lambda$ (which is parity even) and introduces the conjugate twistor variable $\omega$ as 
\begin{align}
p_{\alpha\dot\alpha}&=\lambda_\alpha\bar\lambda_{\dot\alpha} \;,\nn\\
\omega^{\dot\alpha}&=x^{\dot\alpha\alpha}\lambda_\alpha\;.
\label{ParticleTwistorEq}
\end{align}
These relations are written using the $\Spin(1,\nu+1)\simeq\SL(2,\KK_\nu)$ language of Appendix \ref{DivisionAlgApp}, which is convenient for displaying the constraints and local symmetries in the twistor picture.
They are equivalent to $p_a=\tfrac12(\lambda\gamma_a\lambda)$, $\omega^\alpha=x^a(\gamma_a\lambda)^\alpha$, which we will choose to use in $D=10$.
We can indeed check the symplectomorphism property of the twistor transform using eq. \eqref{ParticleTwistorEq}:
\begin{align}
\tfrac12\Re(dp_{\alpha\dot\alpha}\wedge dx^{\dot\alpha\alpha})
&=\tfrac12\Re\left((d\lambda_\alpha\bar\lambda_{\dot\alpha}+\lambda_\alpha d\bar\lambda_{\dot\alpha})\wedge dx^{\dot\alpha\alpha}\right)
=-\Re(\lambda^\dagger dx\wedge d\lambda)\;,\nn\\
\Re(d\lambda_\alpha\wedge d\bar\omega^\alpha)
&=\Re\left(d\lambda_\alpha\wedge(d\bar\lambda_{\dot\alpha}x^{\dot\alpha\alpha} + \bar\lambda_{\dot\alpha}dx^{\dot\alpha\alpha})\right)
=-\Re(\lambda^\dagger dx\wedge d\lambda)
\end{align}
(the two terms on the first row are equal, the first term on the second row vanishes). 
The Poisson bracket on the twistor phase space thus is\footnote{Here, we have switched to real spinor indices, taking $2\nu$ values.}
$\{\lambda_\alpha,\omega^\beta\}=\delta_\alpha^\beta$.

The twistor transform implies the incidence relation (again, $\nu\neq8$)
\begin{align}
T\equiv\Im(\lambda^\dagger\omega)=0\;.
\label{MasslessIncidenceEq}
    \end{align}
This is an element in $\KK'_\nu$, the imaginary elements in $\KK_\nu$ (empty for $\nu=1$). It generates the invariance of $p$ under 
$\lambda\rightarrow\lambda e$, where $e$ is a unit element in $\KK_\nu$, $e\bar e=1$. This R-symmetry 
$S^{\nu-1}\simeq\{\ZZ_2,\U(1),\SU(2)\}$ is the gauge symmetry in the twistor picture and expresses a projective spinor $S^{2\nu-1}$ as the Hopf fibration of $S^{\nu-1}$ over the base $S^\nu$ (the celestial sphere).
The Poisson bracket between constraints when $\nu=4$ is $[T_i,T_j]=2\epsilon_{ijk}T_k$.

The twistor formulation of the massless particle has manifest conformal symmetry $\sp(4,\KK_\nu)\simeq\so(2,\nu+2)$. $\lambda$ and $\omega$ are collected in the 4-component conformal spinor $Z_A$, and the incidence relation is
$\epsilon^{A\dot A}\bar Z_{\dot A}Z^{\mathstrut}_A=0$.

\subsubsection{Infinite spin particles} 

We want to solve $T_1=p\cdot r=0$, alongside with $T_0=p^2=0$. They can both be solved, thanks to the Fierz identity $(\gamma^a\lambda)^\alpha(\lambda\gamma_a\lambda)=0$, 
by
\begin{align}
p_{\alpha\dot\alpha}&=\lambda_\alpha\bar\lambda_{\dot\alpha}\;,\nn\\
r_{\alpha\dot\alpha}&=\lambda_\alpha\bar\nu_{\dot\alpha}+\nu_\alpha\bar\lambda_{\dot\alpha}\;,
\label{InfSpinTwistor1}
\end{align}
with the invariance $\delta\lambda=\lambda a$, $\delta\nu=\nu a+\lambda b$, with $a,b\in\KK'_\nu$.
Let the twistor variable conjugate to $\nu$ be $\chi$.
With the conjugate variables as
\begin{align}
\omega^{\dot\alpha}&=x^{\dot\alpha\alpha}\lambda_\alpha +y^{\dot\alpha\alpha}\nu_\alpha\;,\nn\\
\chi^{\dot\alpha}&=y^{\dot\alpha\alpha}\lambda_\alpha\;,
\label{InfSpinTwistor2}
\end{align}
the incidence relations are:
\begin{align}
&T\equiv\Im(\lambda^\dagger\omega+\nu^\dagger\chi)=0\;,\nn\\
&U\equiv\Im(\lambda^\dagger\chi)=0\;.
\label{InfIncidenceEq}
\end{align}
They generate $S^{\nu-1}\inplus\KK'_\nu$. 
Note that conformal symmetry is broken.

Again, we can show explicitly using the twistor transform, eqs. \eqref{InfSpinTwistor1} and \eqref{InfSpinTwistor2}, that the symplectic reductions on each side of the transform are symplectomorphic:
\begin{align}
&\tfrac12\Re(dp_{\alpha\dot\alpha}\wedge dx^{\dot\alpha\alpha}
+dr_{\alpha\dot\alpha}\wedge dy^{\dot\alpha\alpha})
=\Re(d\lambda_\alpha\wedge d\bar\omega^\alpha
+d\nu_\alpha\wedge d\bar\chi^\alpha)\;.
\end{align}
The remaining constraints for the infinite spin particle are $r^2=1$, $y\cdot p=\mu$.
The second of these takes the form 
$\Re(\lambda^\dagger\chi)=2\mu$.
This constraint combines nicely with the second one in eq. \eqref{InfIncidenceEq} to $\lambda^\dagger\chi=2\mu$.
The constraint $r^2=1$ can be written as
\begin{equation} 1=(\lambda\gamma_a\nu)(\lambda\gamma^a\nu)=
-{1\over2}(\lambda\gamma_a\lambda)(\nu\gamma^a\nu).
\label{rsquared_constr}
\end{equation}
In $D=3,4$ this reads $|\epsilon^{\alpha\beta}\lambda_\alpha\nu_\beta|=1$.
All constraints are first class.

Here, it is appropriate to compare with the description of $D=6$ infinite spin particles by Buchbinder et al. \cite{Buchbinder:2021bgv} using spinorial (twistor-like) variables. The authors check that any value of the (massless) Poincar\'e Casimirs can be obtained. This seems at first sight to be more general than our situation, where the sextic Casimir ($A^2$ in the terminology of eq. \eqref{ABCDdef}) automatically vanishes.
The reason for this is that ref. \cite{Buchbinder:2021bgv} does not start from a space-time action and a twistor transform, but directly postulates dynamics for the spinorial variables.
In particular, incidence relations are {\it not} imposed, neither is the corresponding gauge symmetry. 
We note that the procedure of relaxing the incidence relations closely mirrors that of going from massless particles to higher spin \cite{Cederwall:2004cf}, but in this case from a situation where there is already an infinite tower of higher spin states. 

\subsubsection{Brink--Schwarz superparticle} 
For the Brink--Schwarz superparticle \cite{Brink:1981nb}, the twistor transform 
is extended ($D=3,4,6$) to \cite{Shirafuji:1983zd,Ferber197855,Bengtsson:1987ap,Bengtsson:1987si}
\begin{align}
p_{\alpha\dot\alpha}&=\lambda_\alpha\bar\lambda_{\dot\alpha}\;,\nn\\
\omega^{\dot\alpha}&=x^{\dot\alpha\alpha}\lambda_\alpha +\theta^{\dot\alpha}s\;, \\
s&=\bar\theta^\alpha\lambda_\alpha\;.\nn
\end{align}
The fermionic scalar $s\in\KK_\nu$ contains the physical fermions. The incidence relations are modified to
\begin{align}
T\equiv\Im(\lambda^\dagger\omega)-\bar ss=0\;.
\label{incidence1}
\end{align}
Note that the local $\SU(2)$ generated by the incidence relations transform $\lambda$, $\omega$ and $s$ by right multiplication by unit elements in $\KK_\nu$, while R-symmetry acts only on $s$ by left multiplication.

There is a subtle point in the derivation of the supertwistor symplectic structure. Namely, it should, after symplectic reduction, be equivalent to the Dirac bracket obtained by eliminating the second class part \cite{Bengtsson:1984rw} of the constraint
$\tau_\alpha =\pi_\alpha+p_a(\gamma^a\theta)_\alpha=0$.
This goes as follows.\footnote{Trying to show the equality of the symplectic forms on the surface $\tau_\alpha=0$ as before, by equating a symplectic form on the two sides of the supertwistor transform using constraints, gives the wrong coefficient for $\{s_I,s_J\}$. It is not consistent with the transformation properties of $\lambda$, $\omega$ and $s$. We do not know why this erroneous procedure produces an ``almost correct'' result.}
Choose a frame such that the light-like momentum is 
\begin{align}
p_{\alpha\dot\alpha}=\left(\begin{matrix}k&0\\0&0\end{matrix}\right)\;.
\end{align}
Then,
\begin{align}
\lambda_\alpha=\sqrt k\left(\begin{matrix} e\\0\end{matrix}\right)\;,
\end{align}
where $e\in\KK_\nu$, $|e|=1$.
Inspecting the fermionic constraints
$\tau_{\alpha}=\pi_{\alpha}+p_{\alpha\dot\alpha}\theta^{\dot\alpha}=0$, $\tau_2$ is first class and can be used to eliminate $\theta^2$ and $\pi_2$, while $\tau_1$ is second class. The matrix of Poisson brackets between second class constraints is
\begin{align}
M_{IJ}=\{(\tau_1)_I,(\tau_1)_J\}=\{\Re(\bar e_I\tau_1),\Re(\bar e_J\tau_1)\}
=2k\delta_{IJ}\;,
\end{align}
and we obtain the Dirac brackets\footnote{We take the liberty to sometimes use the same type index $I,J,\ldots$ to denote the real components of an element in $\KK_\nu$, disregarding its transformation properties.}
\begin{align}
\{(\theta^1)_I,(\theta^1)_J\}^\star
=-\{(\theta^1)_I,(\tau_1)_K\}
(M^{-1})^{KL}\{(\tau_1)_L,(\theta_1)_J\}
=-\tfrac1{2k}\delta_{IJ}\;.
\end{align}
Now, $s=\bar\theta^\alpha\lambda_\alpha
=\sqrt k\bar\theta^1e$,
and we obtain (dropping the star on the bracket) 
\begin{align}    
\{s_I,s_J\}=-\tfrac12\delta_{IJ}\;,
\end{align}
independent of choice of frame.

Note that this bracket is consistent with the $\su(2)$ algebra of constraints in $D=6$, and also with $\lambda$ and $s$ transforming with the same $\U(1)$ charge in $D=4$ (which is necessary considering the transformation of $\omega$).

\subsubsection{Infinite spin superparticle} 

As we have seen, the fermionic constraints of the infinite spin superparticle (after some gauge fixing) are identical to those of the Brink--Schwarz superparticle. 
We can then perform a supertwistor transform as a synthesis of those for the bosonic infinite spin particle and for the Brink--Schwarz superparticle:
\begin{align}
    p_{\alpha\dot\alpha} &= \lambda_{\alpha}\bar\lambda_{\dot\alpha}\;,\notag\\
    r_{\alpha\dot\alpha} &= \lambda_\alpha \bar\nu_{\dot\alpha} + \nu_{\alpha}\bar\lambda_{\dot\alpha}\;, \notag\\
    \omega^{\dot\alpha} &= x^{\dot\alpha\alpha} \lambda_{\alpha} + y^{\dot\alpha\alpha}\nu_{\alpha} + \theta^{\dot\alpha}s\;,\notag \\
    \chi^{\dot\alpha} &= y^{\dot\alpha\alpha}\lambda_\alpha\;, \notag \\
    s &=  \bar\theta^{\alpha} \lambda_\alpha\;.
\end{align}

The incidence relations and other constraints are:
\begin{align}
T\equiv\Im(\lambda^\dagger\omega+\nu^\dagger\chi)-\bar ss&=0\;,\nn\\
\lambda^\dagger\chi&=2\mu\;,\\
   (\lambda\gamma_a\nu)(\lambda\gamma^a\nu)&=1\;,\nn
\end{align}

\subsection{10 dimensions}

\subsubsection{Bosonic particle}

For $D=10$ \cite{Berkovits:1990yc,Cederwall:1992bi,cederwall2025octonionicphasespacehopf},
$S^{\nu-1}=S^7$ is not a Lie algebra. Transformation in $S^7$ can nevertheless be defined \cite{Cederwall:1993nx}. The incidence relation \eqref{MasslessIncidenceEq} does not make sense.
Instead, it can effectively be ``multiplied by a spinor'' to obtain
\begin{align}
    T_{\alpha}=-\tfrac12(\lambda\gamma_a\lambda)(\gamma^a\omega)_\alpha+\lambda_\alpha(\lambda\omega)=0 \;.
    \label{S7GenEq}
\end{align}
In 2-component notation the constraint reads
$T_\alpha=(\lambda_\alpha\bar\lambda_{\dot\alpha})\omega^{\dot\alpha}-\lambda_\alpha\mathrm{Re}(\bar\lambda_{\dot\alpha}\omega^{\dot\alpha})=0$ (notice the need of parentheses; when $\KK_\nu$ is associative this expression equals
$\lambda_\alpha \Im(\lambda^\dagger\omega)$).
The Tate resolution turns out to be finite \cite{Berkovits:1990yc}.
The constraint has only a two-step reducibility, consisting of a vector and a scalar:
\begin{align}
T'^a&\equiv(\lambda\gamma^aT)=0\;,\nn\\
T''&\equiv(\lambda\gamma_a\lambda)T'^a=0
\;.
\end{align}
Counting the number of independent constraints, $16-10+1=7$, and they can be shown to generate $S^7$ (see for example \cite{Cederwall:1993nx}).
The Poisson bracket between the constraints is
\begin{align}
\{T_\alpha,T_\beta\}=\lambda_\alpha T_\beta-\lambda_\beta T_\alpha\;.
\end{align}
This resolution of the octonionic Hopf map was given in ref. \cite{Cederwall:1992bi}
and rediscovered in component form in ref. \cite{Nahari:2024}.
The constraints can also be cast in a manifestly conformal form,
$(Z\Gamma^{MN}Z)(\Gamma_{MN}Z)_A=0$, but the corresponding reducibility in terms of an $\so(2,10)$-covariant Tate resolution is not known explicitly. An additive resolution is straightforward
\cite{cederwall2025octonionicphasespacehopf}.
The constraint surface is the 25-dimensional spinor orbit\footnote{This is the minimal non-trivial real spinor orbit; the 16-dimensional minimal orbit is complex.} of $\Spin(2,10)$.

The $\so(1,9)$-invariant BRST operator expressing the resolution is, with ghosts $c^\alpha$, $c'_a$ and $c''$ of ghost numbers 1, 2 and 3 respectively,
\begin{align}
q&=-\Bigl(\tfrac12(\lambda\gamma_a\lambda)(\gamma^ac)_\alpha-\lambda_\alpha(\lambda c)\Bigr)\omega^\alpha
+\Bigl(-(\lambda c)c^\alpha+(\gamma_a\lambda)^\alpha c'^a\Bigr)b_\alpha
\nn\\
&\qquad+\Bigl(-(\lambda\gamma^a\gamma_bc)c'^b+(\lambda\gamma^a\lambda)c''\Bigr) b'_a
+\Bigl(-2(\lambda c)c''+c'^ac'_a\Bigr)b''
\\
&=(cT)+c'^a\tilde T'_a+c''\tilde T''
-(\lambda c)(cb)-(\lambda\gamma^a\gamma_bc)c'^bb_a
-2(\lambda c)c''b''+c'^ac'_ab''
\;,\nn
\end{align}
where $\{b_\alpha,c^\beta\}=\delta_\alpha^\beta$, $\{b'_a,c'^b\}=\delta_a^b$, $\{b'',c''\}=1$, and where
$\tilde T'_a=(\lambda\gamma_a b)$, $\tilde T''=(\lambda\gamma^a\lambda)b'_a$.
The BRST operator of course carries ghost number 1, and also respects a grading, where $\lambda$, $c$, $c'$ and $c''$ carry degrees 1, $-1$, $-2$ and $-4$, respectively (reflecting the degrees 1, 2 and 4 of $T$, $T'$ and $T''$).

\subsubsection{Brink--Schwarz superparticle}

In lower dimension, the way to deal with the Grassmann coordinates was by means of a scalar. In $D=10$ \cite{Berkovits:1990yc,Cederwall:1992bi}, there are no scalar octonions. Instead, if one chooses the other chirality for $\lambda$, the fermions can be assigned to a vector $s^a=(\lambda\gamma^a\theta)$, which now is subject to a constraint 
\[t\equiv(\lambda\gamma_a\lambda)s^a=0\;.\] 

Like for the lower-dimensional cases, the correct bracket between fermions requires checking the Dirac bracket obtained when eliminating the second class constraints. In this case, due to the constraint $t$, it suffices to check the Dirac bracket for some choice of gauge. A calculation parallelling the one for lower dimensions gives
\begin{align}
\{s^a,s^b\}=-\tfrac12\eta^{ab}\;,
\end{align}
which of course also is consistent with the algebra of the constraints.

In addition, after the conjugate $\omega$ is defined as
\begin{align}
\omega_\alpha = x^a(\gamma_a\lambda)_\alpha +  s^a (\gamma_a\theta)_\alpha 
\end{align}
the $S^7$ constraint becomes
\begin{align}
    T^{\alpha}=-\tfrac12(\lambda\gamma_a\lambda)(\gamma^a\omega)^\alpha+\lambda^\alpha(\lambda\omega)+s^as^b(\gamma_{ab}\lambda)^\alpha=0 \;.
    \label{S7GenEq2}
\end{align}
The Poisson brackets between the constraints are
\begin{align}
\{T^\alpha,T^\beta\}&=\lambda^\alpha T^\beta-\lambda^\beta T^\alpha\;,\nn\\
\{T_\alpha,t\}&=\lambda^\alpha t\;,\\
\{t,t\}&=0\;.\nn
\end{align}

The reducibility is
\begin{align}
T'^a&\equiv(\lambda\gamma^aT)-2s^at=0\;,\nn\\
T''&\equiv(\lambda\gamma_a\lambda)T'^a=0
\;.
\end{align}
The resolution of the constraints was given, in terms of a BRST operator, in ref. \cite{Cederwall:1992bi}:
\begin{align}
q&=(cT)+\gamma t+c'^a\tilde T'_a+c''\tilde T''\nn\\
&-(\lambda c)(cb)-(\lambda\gamma^a\gamma_bc)c'^bb_a
-2(\lambda c)c''b''+c'^ac'_ab''
-(\lambda c)\gamma\beta
\;,
\end{align}
where now 
$\tilde T'_a=(\lambda\gamma_ab)-2s_a\beta$,
$\tilde T''=(\lambda\gamma^a\lambda)b'_a+\beta^2$
and where $\{\beta,\gamma\}=1$. 
The second term in $\tilde T''$ was unfortunately missing in ref. \cite{Cederwall:1992bi}. The precise statement is that $\tilde T''$, as given here, represents the cohomology trivialised by the introduction of $c''$; this is not seen by only considering sequential reducibility.

\subsubsection{Infinite spin superparticle}

The $D=10$ twistor transform for the supersymmetric infinite spin particle, generalising the one for the Brink--Schwarz superparticle \cite{Berkovits:1990yc,Cederwall:1992bi}, is
\begin{align}
    p_a &= \tfrac12(\lambda\gamma_a\lambda)\;,
    \notag\\
    r_a &= (\lambda\gamma_a\nu) \;,\notag\\
    \omega_\alpha &= x^a(\gamma_a\lambda)_\alpha+ y^a(\gamma_a\nu)_\alpha +  s^a (\gamma_a\theta)_\alpha \;,\label{d10supertwistor}\\
    \chi_\alpha &=  y^a(\gamma_a\lambda)_\alpha\;, \notag \\
    s^a &=  (\lambda\gamma^a \theta)\;.\nn
\end{align}

The constraints following from the super-twistor transform \eqref{d10supertwistor} are
\begin{align}
T^\alpha&\equiv-\tfrac12(\lambda\gamma_a\lambda)(\gamma^a\omega)^\alpha+\lambda^\alpha(\lambda\omega)
+s^as^b(\gamma_{ab}\lambda)^\alpha\nn\\
&\quad-(\lambda\gamma_a\nu)(\gamma^a\chi)^\alpha
+\lambda^\alpha(\nu\chi)+\nu^\alpha(\lambda\chi)
=0\nn\;,\\
U^\alpha&\equiv-\tfrac12(\lambda\gamma_a\lambda)(\gamma^a\chi)^\alpha+\lambda^\alpha(\lambda\chi)
=0\;,\\
t&\equiv(\lambda\gamma_a\lambda)s^a=0\;,\nn
\end{align}
with the reducibility
\begin{align}
T'^a&\equiv(\lambda\gamma^aT)+(\nu\gamma^aU)+2s^at=0\;,\nn\\
U'^a&\equiv(\lambda\gamma^aU)=0\;,
\end{align}
and second order reducibility
\begin{align}
T''&\equiv(\lambda\gamma_a\lambda)T'^a
+2(\lambda\gamma_a\nu)U'^a=0\;,\nn\\
U''&=(\lambda\gamma_a\lambda)U'^a=0\;.
\end{align}
In addition, we still have the constraints
$H=\tfrac12(\lambda\chi)-\mu=0$ and $T_2=(\lambda\gamma_a\nu)(\lambda\gamma^a\nu)-1=0$.
The non-vanishing Poisson brackets between constraints are:
\begin{align}
    \{T^\alpha, T^\beta\}&=\lambda^\alpha T^\beta -\lambda^\beta T^\alpha \;,\nn\\
    \{T^\alpha, U^\beta\}&=\lambda^\alpha U^\beta - \lambda^\beta U^\alpha \;,\nn\\
    \{T^\alpha,t\}&=\lambda^\alpha t\;,\\
    \{H,T^\alpha\}&=-\tfrac12U^\alpha\;.\nn
\end{align}
    
This allows us to write down a BRST operator,
\begin{align}
q&=(cT)+c'^a\tilde T'_a+c''\tilde T''+\gamma t\nn\\
&+(gU)+g'^a\tilde U'_a+g''\tilde U''\nn\\
&+c_HH+c_2T_2+\tfrac12 c_H(cf)\\
&-(\lambda c)(cb)-(\lambda\gamma^a\gamma_bc)c'^bb_a
-2(\lambda c)c''b''+c'^ac'_ab''
-(\lambda c)\gamma\beta\nn\\
&-(\lambda c)(gf)-(\lambda g)(cf)
-2(\lambda c)g''f''-2(\lambda g)c''f''+2c'^ag'_af''\nn
\;,
\end{align}
where 
\begin{align}
\tilde T'^a&=(\lambda\gamma^a b)+(\nu\gamma ^a f)+2s^a\gamma\;,\nn\\
\tilde U'^a&=(\lambda\gamma^af)\;,\nn\\
\tilde T''&=(\lambda\gamma_a\lambda)b'^a
+2(\lambda\gamma_a\nu)f'^a +\beta^2   \;,\\
\tilde U''&=(\lambda\gamma_a\lambda)f'^a    \;,\nn
\end{align}
where the ghosts for $U$
and its reducibilities are denoted $f,g$ in analogy to $b,c$, and the ghosts for $H$ and $T_2$ are $b_H,c_H$ and $b_2,c_2$.

\section{Quantisation}

\subsection{Bosonic infinite spin particle}

Consider a bosonic infinite spin particle, with phase space variables $\{x,p;y,r\}$ and the first class constraints
\begin{align}
T_0&\equiv p^2\approx0\;,\nn\\
T_1&\equiv p\cdot r\approx0\;,\nn\\
T_2&\equiv r^2-1\approx0\;,\\
H&\equiv p\cdot y-\mu\approx0\;.\nn
\end{align}
Let us choose a polarisation where the one-particle states are described as functions of $p$ and $r$. Then, $y^a=i{\*\over\*r_a}$. The polarisation is convenient, since then $T_{0,1,2}$ are algebraic constraints on the variables, not differential equations.
Given a light-like $p$, $r$ can be parametrised as
$r=\varrho+tp$, where $\varrho$ parametrises (non-canonically) an $S^{d-3}$ inside the $(d-2)$-dimensional hyperboloid of unit vectors orthogonal to $p$, 
$\varrho^2=1$, $\varrho\cdot p=0$. For example, in a frame where
$p=k(1,0,\ldots,0,1)$, we can choose $\varrho=(0,\vec\varrho,0)$.
Then, $p\cdot{\*\over\*r}={\*\over\*t}$, and
the Hamiltonian constraint reads ${\*\over\*t}-i\mu=0$.
(${\*\over\*t}$ is scalar and independent of the choice of $\varrho$.)
A momentum mode of the wave function is
\begin{align}
\psi_p(r)=e^{i\mu t}f_p(\vec\varrho) \;.
\end{align}
The functions $f_p$ can be expanded in spherical harmonics on $S^{d-3}$.
Spherical harmonics (traceless powers of $\vec\varrho$) come in $\so(d-2)$ modules described by completely symmetric and traceless tensors, with Dynkin indices
$(j0\ldots0)$. 
This gives the content of the infinite spin model as the sum of all spins
described by Fronsdal fields \cite{Bekaert:2017khg}.

The quantisation is not unique. One may, as above, choose the algebra of spherical harmonics as the Hilbert space for the $r$
variables, but any module of this algebra serves as well. For example, in $D=4$, bosonic massless fields of integer spins are obtained as the functions $z^n$, $n\in\ZZ$, where $z=\varrho_1+i\varrho_2$.
An alternative quantisation of the same bosonic infinite spin particle, describing (fermionic) massless fields with spins in $\ZZ+\frac12$, is described by the module consisting of functions on the double cover, $z^r$, $r\in\ZZ+\frac12$.
In the following subsections, we will show how these two modules combine in the supersymmetric infinite spin particle, and also perform quantisation of the supersymmetric infinite spin particle in higher dimensions. 

\subsection{Supersymmetric infinite spin particles}

We note that the difference between the Brink--Schwarz superparticle and the infinite spin superparticle is (maybe surprisingly) small. The fermionic variables, after some gauge fixing, behave exactly as in the Brink--Schwarz situation.
The variable $s$ (for $D\neq10$) transforms under the gauge $S^{\nu-1}$ but commutes with all the ``extra'' variables $\nu$, $\chi$ giving the infinite tower of higher spin states.    

The constraints we have to respect are:
\begin{align}
    \bar\lambda \omega -\bar \omega \lambda + \bar \nu \chi - \bar \chi\nu -2 \bar s s &=0\;,\\
     (\lambda \gamma^a \nu) (\lambda \gamma_a \nu) - 1&=0\;,\\
     \bar\lambda \chi - \mu &=0 \;.\label{Ham:true-eq}
\end{align}
The first one imposes to work with gauge singlets, the second tells us about tracelessness of the tensors (spherical harmonics, for a given light-like $p$) and it is also an algebraic condition that defines a variety, while the third gives a differential equation relating different tensors (the real part) and a gauge symmetry (the imaginary part).

In the twistor space we choose the polarization or Lagrangian subspace with \emph{bosonic} coordinates
$\lambda_\alpha$, $\nu_\alpha$.
We use, in this Section, the expressions $p_a$ and $r_a$ for the gauge invariants $\tfrac12(\lambda\gamma_a\lambda)$ and 
$(\lambda\gamma_a\nu)$.

The fermionic variables $s$, on the other hand, 
transform under gauge $S^{\nu-1}$ and also under $R$-symmetry.
The real components of $s$ are self-conjugate.
One may apply ``$\gamma$-matrix quantisation'', treating the components of $s$ as $\gamma$ matrices, acting on a Dirac spinor of $\Spin(\nu)$. This fermionic Hilbert space will however not consist of gauge-invariant states. See below how this can be remedied in $D=6$. A better strategy is in general to construct the gauge-invariant operator $q_\alpha=\lambda_\alpha\bar s$, which is the on-shell supersymmetry generator, and, let it generate the physical states.

\subsubsection{4 dimensions}

We find a tower of fermions and one of bosons: these are generated by the gauge invariant operators $q, \bar q$ and can also be found by tensor product multiplication of symmetric traceless tensors (functions of $r$).

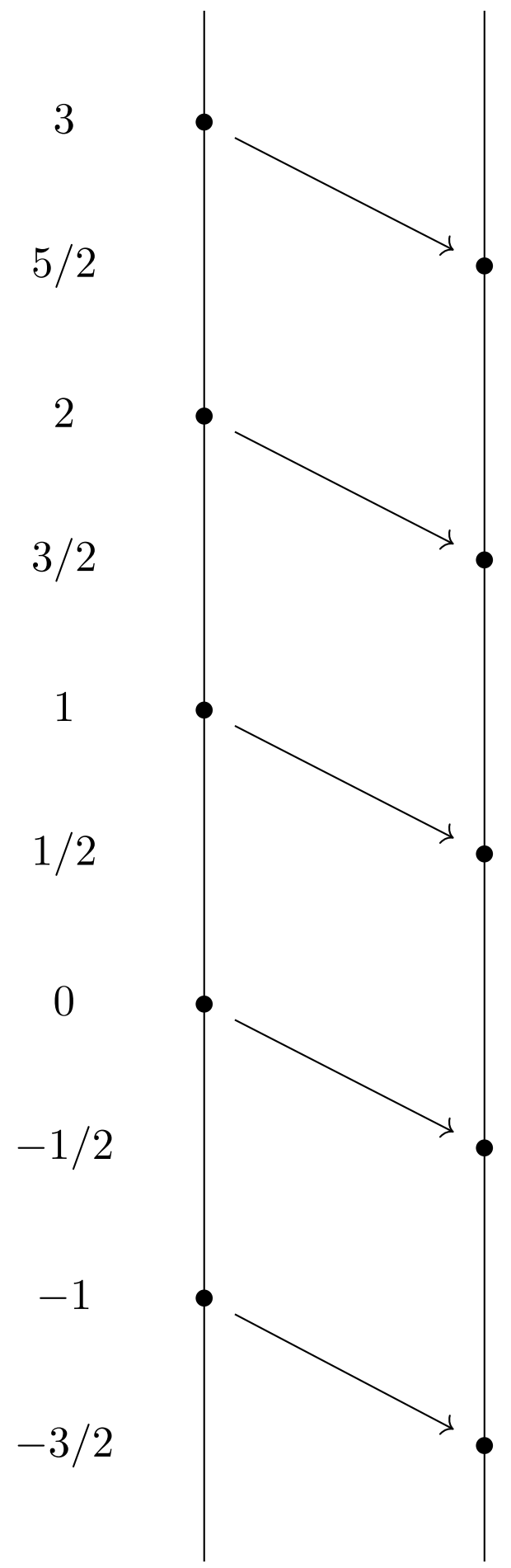
\begin{figure}
\begin{center}
\begin{tikzcd}[column sep={32pt,between origins}]
&\phantom{}\ar[ddddddddddd,no head]
    &&\phantom{}\ar[ddddddddddd,no head]
\\
3&\bullet\ar[drr]\\
5/2&&&\bullet\\
2&\bullet\ar[drr]\\
3/2&&&\bullet\\
1&\bullet\ar[drr]\\
1/2&&&\bullet\\
0&\bullet\ar[drr]\\
-1/2&&&\bullet\\
-1&\bullet\ar[drr]\\
-3/2&&&\bullet\\
&\phantom{}&&\phantom{}
\end{tikzcd}
\caption{The supersymmetric combination of a bosonic tower of states and a fermionic one in $D=4$. The labels on the vertical axis are helicities (transverse $U(1)$ charge).\label{TableD4MinSpectrum}}
\end{center}
\end{figure}

\begin{figure}
\begin{center}
\begin{tikzcd}[column sep={32pt,between origins}]
&\phantom{}\ar[dddddddddddd,no head]
    &&\phantom{}\ar[dddddddddddd,no head]
    &&\phantom{}\ar[dddddddddddd,no head]
    &&\phantom{}\ar[dddddddddddd,no head]\\
3&\bullet\ar[drr]\ar[drrrr,blue,dashed]&&&&&&\bullet\ar[dllll,magenta,dotted]\\
5/2&&&\bullet\ar[drrrr,blue,dashed]&&\bullet\ar[drr]\ar[dllll,magenta,dotted]\\
2&\bullet\ar[drr]\ar[drrrr,blue,dashed]&&&&&&\bullet\ar[dllll,magenta,dotted]\\
3/2&&&\bullet\ar[drrrr,blue,dashed]&&\bullet\ar[drr]\ar[dllll,magenta,dotted]\\
1&\bullet\ar[drr]\ar[drrrr,blue,dashed]&&&&&&\bullet\ar[dllll,magenta,dotted]\\
1/2&&&\bullet\ar[drrrr,blue,dashed]&&\bullet\ar[drr]\ar[dllll,magenta,dotted]\\
0&\bullet\ar[drr]\ar[drrrr,blue,dashed]&&&&&&\bullet\ar[dllll,magenta,dotted]\\
-1/2&&&\bullet\ar[drrrr,blue,dashed]&&\bullet\ar[drr]\ar[dllll,magenta,dotted]\\
-1&\bullet\ar[drr]\ar[drrrr,blue,dashed]&&&&&&\bullet\ar[dllll,magenta,dotted]\\
-3/2&&&\bullet\ar[drrrr,blue,dashed]&&\bullet\ar[drr]\ar[dllll,magenta,dotted]\\
-2&\bullet&&&&&&\bullet\\
&\phantom{}&&\phantom{}&&\phantom{}&&\phantom{}
\end{tikzcd}
\caption{The double tower of states of the $D=4$ infinite spin superparticle. The solid arrows show the $N=1$ supersymmetry generator. The doubling of states makes it possible to obtain supermultiplets symmetric around helicity 0. 
$N=2$ supersymmetry can be realised on the doubled Hilbert space.
The dashed arrows indicate the action of a second supersymmetry generator when the $r$-independent part of the spectrum is the $N=2 $ Maxwell multiplet, the dotted ones when it is the hypermultiplet. Only in the former case, the supermultiplets have highest helicity components at even helicity. \label{TableD4Spectrum}}
\end{center}
\end{figure}
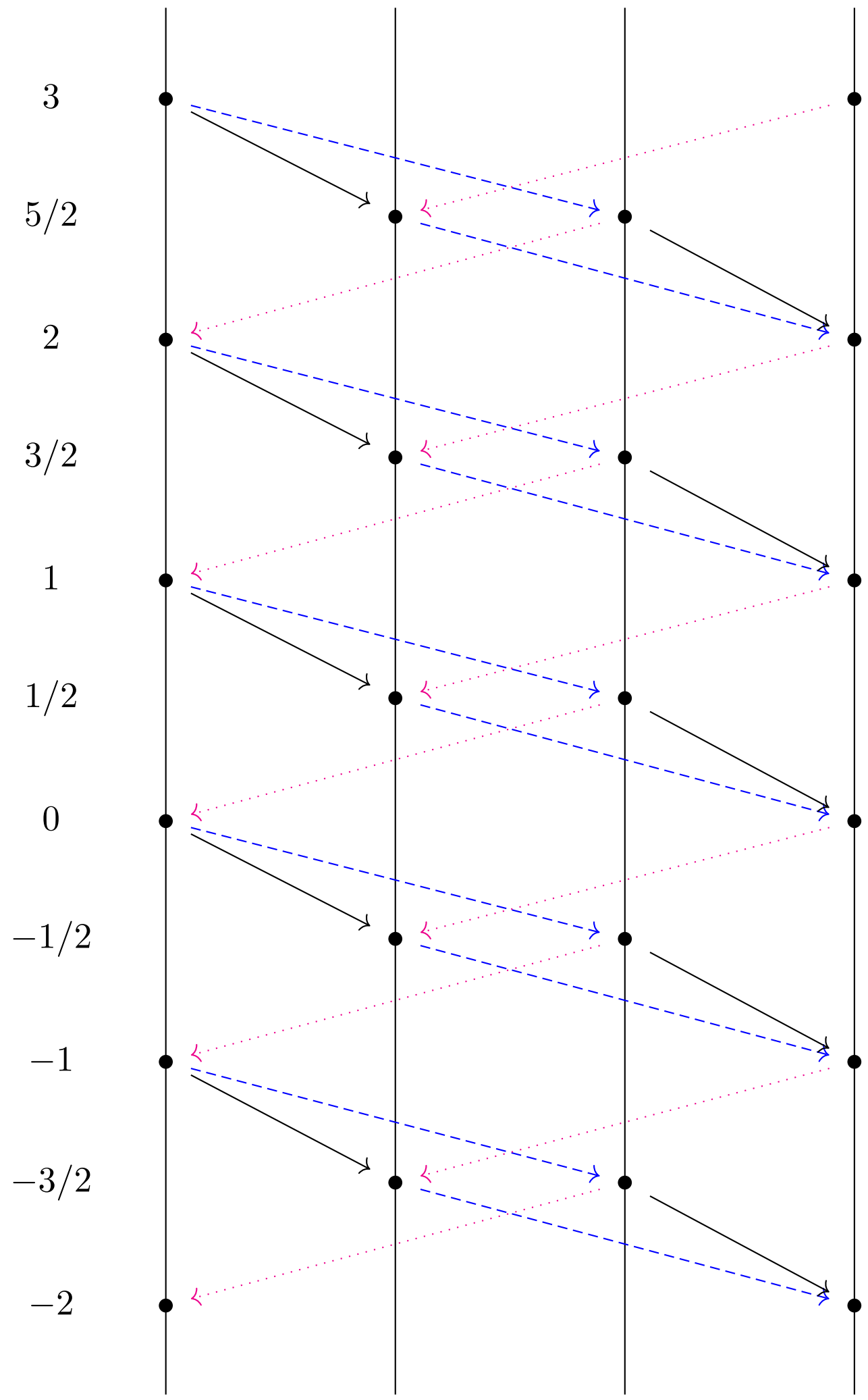

This ``minimal'' representation is depicted in Figure \ref{TableD4MinSpectrum}. We note that although the supersymmetry generator commutes with $r$, each ``spherical harmonic'' is associated with a half supermultiplet, without its CPT conjugate. We can remedy this by adding a second copy of integer and half-integer states as in Figure \ref{TableD4Spectrum}.
Then, $N=2$ supersymmetry may be realised. Depending on how this is done, the spectrum either consists of shifted copies of a hypermultiplet or of a Maxwell multiplet. The latter option, having integer highest helicity, seems sounder.

\subsubsection{6 dimensions}

Even though we described above how to obtain states from gauge-invariant operators,
suppose now we perform a $\gamma$-matrix quantisation for $s$. We then obtain states in chiral spinors $\phi$ and $\eta$, transforming under $\su(2)$ R-symmetry and $\su(2)$ gauge symmetry, respectively. (This holds for the case at hand as well as the Brink--Schwarz superparticle.)
We can use conventions where these $\su(2)$'s act by left multiplication on the two spinors, and a new R' $\su(2)$ acts from the right (it has to be the same for the two spinors in order for a vector to map from one chirality to the other).
However, we need to consider gauge-invariant states, so a naked $\eta$ is not allowed. Forming 
$\psi_\alpha=\lambda_\alpha\eta$, we have constructed a spinor of gauge-invariant states. 
Note that this spinor satisfies the Dirac equation, thanks to the Fierz identity $(\gamma_a\lambda)^\alpha(\lambda\gamma^a\lambda)=0$.
So, quantisation of the fermions has led to states labelled by their transformation under Lorentz $D_3$, R-symmetry and R'-symmetry, specified by Dynkin labels:
\begin{align}
\phi&:\quad\quad\DWeight000(1)(1)\;,\nn\\
\psi&:\quad\quad\DWeight010(0)(1)\;.
\end{align}
This is an on-shell hypermultiplet of $N=1$ supersymmetry in $D=6$.

This is of course a natural result in the quantisation of the Brink--Schwarz superparticle. (Also any other supermultiplet can be obtained.)
But in the context of infinite spin, it is not desirable.
Choosing this quantisation will not give any bosonic gauge fields (Maxwell, gravity,...) which are also scalar under R-symmetry.
This gives an infinite spin representation of the super-Poincar\'e algebra, but it is not physically appealing.

As described above,
the (more general) alternative to the above $\gamma$-matrix quantisation is to construct gauge-invariant {\it operators} instead of states. Such operators can then generate states from any vacuum. So, instead of letting the components of $s$  
(the $\gamma$-matrices) generate gauge-non-invariant states, we consider the gauge-invariant operator $q_\alpha=\lambda_\alpha\bar s=p_{\alpha\dot\alpha}\theta^{\dot\alpha}$.
This operator transforms under R-symmetry from the right, and is indeed the on-shell supersymmetry generator:
\begin{align}
\{q_\alpha,q_\beta\}
=-\tfrac12(\lambda\bar e_I)_\alpha(\lambda\bar e_I)_\beta
=-\tfrac14\gamma^a_{\alpha\beta}(\lambda\gamma_a\lambda)\;,
\end{align}
where we have used the identities
\begin{align}
\bar e_I\gamma^a e_I&=4\gamma^a\;,\nn\\
\bar e_I\gamma^{abc}e_i e_I&=0
\end{align}
for the two irreducible $\so(1,5)\oplus\su(2)$ modules in the symmetric product of a chiral spinor.

With this approach, using the gauge invariant operator $q_\alpha$ instead of the ``spinor'' on which $s$ acts as gamma matrices, we see that we are directly led to any on-shell massless supermultiplet in the twistor quantisation of the 
Brink--Schwarz superparticle.

An alternative choice for the $r$-independent component of the wave function for the infinite spin superparticle may be
the tensor multiplet, obtained by acting with $q$ on a single scalar. The multiplet contains a scalar, a chiral spinor and a chiral 2-form. On-shell, it can be seen as a reinterpretation/twisting of the hypermultiplet, where only the diagonal subalgebra of the two R-symmetry $\su(2)$'s is kept, and identified with an $\su(2)$ subalgebra of transverse spatial rotations. 

The most appealing option seems to be to start from a vector multiplet.
Let us calculate higher powers of $r$ by tensor product and identify states.
The on-shell vector multiplet, in terms of 
$\su(2)\oplus\su(2)\oplus\su(2)$ modules (transverse $\so(4)$ and R-symmetry $\su(2)$) is
\begin{align}
{\frak m}_0=(1)(1)(0)\ominus(0)(1)(1)\;,
\end{align}
where the minus sign is reserved for fermionic states. 
We take the tensor product with the spherical harmonics in $(n)(n)(0)$ to obtain
\begin{align}
{\frak m}_n&=(n)(n)(0)\otimes{\frak m}_0\nn\\
&=\left[(n+1)(n+1)(0)\ominus(n)(n+1)(1)\oplus(n-1)(n+1)(0)\right]
\label{d6states}\\
&\oplus\left[(n+1)(n-1)(0)\ominus(n)(n-1)(1)\oplus(n-1)(n-1)(0)\right]\;.\nn
\end{align}
At each $n\geq1$, two irreducible massless supermultiplets are produced, the two lines of eq. \eqref{d6states}. At $n=1$, we find the supergravity multiplet, consisting of the metric, gravitino and self-dual 2-form, together with the tensor multiplet.
It is not possible to remove any of the multiplets. Ideals are absent because of the condition $r^2=1$; multiplication with $r$ results both in higher and lower spherical harmonics (unlike the situation for a ring where multiplication preserves a grading).

\subsubsection{10 dimensions}

The BRST operator in the twistor picture looks complicated. Its purpose however, is essentially to yield functions of $p$ and $r$, constrained as before, and to trivialise unwanted cohomology related to reducibility.
Analogously to what happens in lower dimensions, physical states will be generated by the on-shell supersymmetry generator, which now is $q_\alpha=2s^a(\gamma_a\lambda)_\alpha =-2p^a(\gamma_a\theta)_\alpha$. 
Since $q_\alpha$ is expressible in the super-space-time variables, it is gauge invariant. Direct calculation gives
\begin{align}
[q_\alpha,q_\beta]&=-2(\gamma_b\lambda)_\alpha(\gamma^b\lambda)_\beta
=-\tfrac18\gamma^a_{\alpha\beta}(\lambda\gamma^b\gamma_a\gamma_b\lambda)
=\gamma^a_{\alpha\beta}(\lambda\gamma_a\lambda)=2\gamma^a_{\alpha\beta}p_a\;,
\end{align}
using $\gamma^i\gamma_{abcde}\gamma_i=0$.

The most reasonable choice for the $r$-independent part is the smallest supermultiplet, the super-Yang--Mills multiplet. On shell it is given by the transverse $\so(8)$ module 
\begin{align}
{\frak m}_0=\DWeight{10}00\ominus\DWeight{00}10\;,
\end{align}
where the minus sign is reserved for fermions.
Higher $S^7$ spherical harmonics give the tensor products 
${\frak m}_n=\DWeight{n0}00\otimes{\frak m}_0$, $n\geq1$,
\begin{align}
{\frak m}_n
=\DWeight{n+1,0}00\oplus\DWeight{n-1,1}00
    \oplus\DWeight{n-1,0}00
\ominus\DWeight{n0}10\ominus\DWeight{n-1,0}01\;.
\end{align}
These are irreducible $N=1$ supermultiplets with highest spin $n+1$. For $n=1$, we recognise the on-shell supergravity multiplet 
expressed in terms of $\so(8)$ modules,
containing the metric, $B$-field and dilaton, together with a gravitino and a spinor.

\section{Conclusions}

We have formulated the dynamics of (target space) supersymmetric particles in 3, 4, 6 and 10 dimensions through a generalisation of the higher-derivative action (quadratic in acceleration) of ref. \cite{Edgren:2005gq}.
In the process, new results concerning massless super-Poincar\'e Casimir operators were obtained. 
In order to circumvent the need to eliminate half-spinors of fermionic second-class constraints, we also generalised super-twistor transforms to infinite spin. 

Finally, we performed quantisation, yielding a Hilbert space of 1-particle states for infinite spin field theory. We privileged quantisations that yielded gauge multiplets, or that had integer spins as highest weights.
We do not suggest a field theory giving the same on-shell states. Such a theory requires extra bosonic variables. 
We refer to the actions of refs.
\cite{Schuster:2013pta,Bekaert:2015qkt,Najafizadeh:2019mun}, with unclear geometric origins.

The present particle formulation makes the results
\cite{Font:2013hia,Alabbasi:2026kbf} concerning the absence of infinite spin representations in string and superstring theory natural. It also points towards a natural conjecture how a string theory should be modified in order to contain such states, namely by higher world-sheet derivatives.

We have seen that our infinite spin superparticle has a non-zero value of a quartic super-Poincar\'e Casimir operator. In $D=6$, we have demonstrated the existence of a sextic Casimir, which vanishes in our model. We were not able to find further invariants in $D=10$.
Buchbinder et al. \cite{Buchbinder:2021bgv}
obtain non-vanishing values also of a sextic Casimir in a twistor-like model for bosonic infinite spin particles in $D=6$. In our formalism, the price paid is the relaxation of the twistor incidence relations.
Our approach can suggest that representations with non-vanishing values of higher Casimirs {\it might} arise from particle actions with even higher number of world-line derivatives. 

We would finally like to comment on the possible ways of formulating a supersymmetric infinite spin field theory.
It has gradually become clear that supersymmetric field theory is best formulated in the pure spinor superfield framework,\cite{Cederwall:2013vba,Cederwall:2022fwu,Cederwall:2023wxc,Eager:2021wpi} and refs. therein.
We find it plausible that this formalism may be generalised to infinite spin. Some questions remain to be solved, however.
We saw in the present paper that the transversality of the extra bosonic variables, $p\cdot r=0$, was naturally solved by extending the concept of twistors. In the pure spinor superfield formalism, we do not immediately see where it enters. In fact, even $p^2=0$ is not an input, but a consequence of the solution to a cohomological problem. It seems likely that also $p\cdot r=0$ should be incorporated in a similar way. We do not have a solution to this problem at present. However, it can be noted that there exists, in $D=6$, a sequence of $N=1$ supersymmetric off-shell multiplets described by pure spinor superfields with $n$ traceless vector indices, where the maximal spin is $n+1$. An infinite collection of 
such fields may provide a starting point.

\subsection*{Acknowledgements}

The authors express thanks to
Alex Arvanitakis, Evgeny Skvortsov and Vasileios Letsios for kindly answering questions and sharing insights concerning their work.
We acknowledge the COST action CA21109 CaLISTA supported by COST (European Cooperation in Science and Technology) for an STSM that allowed us to conduct some of this research work face-to-face. E.B. thanks Chalmers University for hospitality. M.C. does the same to LMU, Munich, and to B. Jur\v co at Charles Univ., Prague. 
The work of E.B. is supported by GA\v{C}R grant PIF-OUT 24-10634O.   

\newpage

\begin{appendix}

\section{Ostrogradsky formalism on a $\ZZ_2$-graded space\label{OstroApp}}

This appendix recalls and sets the conventions for Ostrogradsky's Hamiltonian formalism \cite{Ostrogradsky:1850fid}
for a Lagrangian depending on position, velocity and acceleration.

Starting from a Lagrangian $L(x,\dot x,\ddot x)$ for a collection of dynamical variables $x^i$ (possibly graded, so we keep track of order and signs), 
a second phase space ``coordinate'' can be taken as $y^i=\dot x^i$.
The momenta conjugate to $x^i$ and $y^i$ are denoted $p_i$ and $r_i$, respectively.
\begin{align}
p_i&={\*L\over\*\dot x^i}-{d\over dt}{\*L\over\*\ddot x^i}\;,\nn\\
r_i&={\*L\over\*\ddot x^i}\;.
\end{align}
Derivatives act from the left unless otherwise indicated.
Perform a Legendre transform to the Hamiltonian:
\begin{align}
H=\dot x^ip_i+\dot y^i r_i-L\;.
\end{align}
Then,
\begin{align}
dH=y^idp_i-dx^i{\*L\over \*x^i}+\dot y^idr_i-dy^i\dot r_i
\end{align}
(this $d$ is here formally taken not to change parity)
from which we read
\begin{align}
H\Cev{\*\over\*p_i}&=y^i(=\dot x^i)\;,\nn\\
{\*H\over\*x^i}&=-{\*L\over\*x^i}(=-\dot p_i)\;,\nn\\
H\Cev{\*\over\*r_i}&=\dot y^i\;,\\
{\*H\over\*y^i}&=-\dot r_i\;.
\end{align}
Using ${\*H\over\*\xi}=(-1)^{|\xi|}H\Cev{\*\over\*\xi}$,
one obtains, for all variables, $\dot\xi=\{H,\xi\}$, where
\begin{align}
&\{A,B\}\nn\\
&=A\left(\Cev{\*\over\*p_i}\Vec{\*\over\*x^i}
            -(-1)^{|x^i|}\Cev{\*\over\*x^i}\Vec{\*\over\*p_i}+\Cev{\*\over\*r_i}\Vec{\*\over\*y^i}
            -(-1)^{|y^i|}\Cev{\*\over\*y^i}\Vec{\*\over\*r_i}\right)B \;.
\end{align}
This Poisson bracket
is graded antisymmetric, $\{B,A\}=-(-1)^{|A||B|}\{B,A\}$, and satisfies a graded Jacobi identity, so it is a graded Lie bracket.
The elementary brackets are $\{p_i,x^j\}=\delta_i{}^j=\{r_i,y^j\}$.

\section{$\SL(2,\KK_\nu)$ and spinors\label{DivisionAlgApp}}

It is useful, in particular in the context of twistors, to use the isomorphisms
$\SL(2,\KK_\nu)\simeq\Spin(1,\nu+1)$, where $\nu=1,2,4,8$ and $\KK_\nu$ is the $\nu$-dimensional division algebra ($\RR$, $\CC$, $\HH$, $\OO$) \cite{Sudbery_1984,Kugo:1982bn,Bengtsson:1987si}.
Chiral ($\nu>1$) spinors
are elements in $\KK_\nu^2$, acted on by sigma matrices which are $2\times2$ Hermitean matrices,
\begin{align}
(\sigma^0)_{\alpha\dot\alpha}
&=\left(\begin{matrix}1&0\\0&1\end{matrix}\right)
=-(\tilde\sigma^0)^{\dot\alpha\alpha}\;,\nn\\
(\sigma^{\nu+1})_{\alpha\dot\alpha}
&=\left(\begin{matrix}1&0\\0&-1\end{matrix}\right)
=(\tilde\sigma^{\nu+1})^{\dot\alpha\alpha}\;,\\
(\sigma^i)_{\alpha\dot\alpha}
&=\left(\begin{matrix}0&\bar e_i\\e_i&0\end{matrix}\right)
=(\tilde\sigma^i)^{\dot\alpha\alpha}\;,\;\;
i=1,\ldots,\nu\;,\nn
\end{align}
where $e_i$ are standard orthogonal basis elements for $\KK_\nu$, and $\bar e_i$ their $\KK_\nu$-conjugates, $\bar e_0=e_0$, $\bar e_j=-e_j$, $j=1,\ldots,\nu-1$. We occasionally define real and imaginary parts of $x\in\KK_\nu$ as 
$\Re x=\tfrac12(x+\bar x)$, $\Im x=\tfrac12(x-\bar x)$.

Vectors are thus represented as Hermitean matrices 
$v_{\alpha\dot\alpha}=v_a(\sigma^a)_{\alpha\dot\alpha}$ or
$\tilde v^{\dot\alpha\alpha}=v_a(\tilde\sigma^a)^{\dot\alpha\alpha}$,
$\tilde v=v-I\,\tr v$. Also,
${1\over2}\Re(v_{\alpha\dot\alpha}\tilde w^{\dot\alpha\alpha})=v\cdot w$.

When $\nu<8$, spinors are also transformed by the R-symmetry group 
$S^{\nu-1}\simeq\{\ZZ_2,\U(1),\SU(2)\}$, elements in $\KK_\nu$ of unit length.
They multiply the spinors from the right (which is important only in six dimensions).

\end{appendix}

\newpage

\bibliographystyle{utphysmod2}
\bibliography{references}

\end{document}